\documentclass{appolb}
\usepackage{epsfig,amssymb,amsmath}

\begin{document}
\title{
%
THE COSMOLOGICAL CONSTANT AND HIGGS MASS \\
WITH EMERGENT GAUGE SYMMETRIES 
}
{\author{Steven D. Bass
\address{Kitzb\"uhel Centre for Physics, 
Hinterstadt 12, 6370 Kitzb\"uhel, Austria}
\address{
Marian Smoluchowski Institute of Physics, 
Jagiellonian University, \\
\L{}ojasiewicza 11,
30-348 Krak\'ow, Poland}
\and
Janina Krzysiak
\address{Institute of Nuclear Physics,
Polish Academy of Sciences, \\
ul. Radzikowskiego 152,
31-342 Krak\'ow, Poland}
}
\maketitle
\begin{abstract}
\noindent
We discuss the Higgs mass and cosmological constant in the
context of an emergent Standard Model,
where the gauge symmetries 
``dissolve'' in the extreme ultraviolet.
In this scenario
the cosmological constant scale is suppressed by power of 
the large scale of emergence and expected to be of similar 
size to neutrino masses.
Cosmology constraints then give an anthropic upper bound on the Higgs mass.
\end{abstract}

\section{Introduction}

The Standard Model provides an excellent description of all
particle physics experiments so far,
from LHC energies to low-energy precision measurements.
The interactions of Standard Model 
particles are determined by gauge symmetries.
Their masses come from coupling
to the scalar Higgs field with non-vanishing vacuum expectation value, vev.
Additional mass is generated in QCD from 
non-perturbative confinement physics with 
dynamical chiral symmetry breaking, 
with about 99\% of the mass of the
hydrogen atom coming from the QCD
confinement potential.
Higgs and QCD condensates fill all space, 
with values independent of the point in free space.

Open puzzles include the origin of 
the gauge symmetries which determine particle dynamics 
and
the hierarchies of scales in particle physics.
The Higgs mass is very much less than the Planck scale
despite a quadratically divergent counterterm which 
naively pushes its value towards the highest scales.
The cosmological constant or vacuum energy density which drives the accelerating expansion of the Universe is characterized by a scale 0.002 eV~\cite{Aghanim:2018eyx}, 
very much less than the QCD, Higgs mass and Planck scales.

Here we argue that
the tiny value of the cosmological constant
may be telling us about 
the deeper origin of gauge symmetries in particle physics
-- that they may be emergent in the infrared, 
``dissolving'' 
in the ultraviolet close to the Planck scale 
(instead of extra unification) \cite{Bass:2020egf}.
Given the incredible success of the Standard Model
with no new particles or interactions seen so far
in our experiments, 
perhaps the symmetries of the Standard Model are more special
than previously anticipated.
The Standard Model with measured parameters works as 
a consistent theory up to the Planck scale with a 
Higgs vacuum that sits very close
to the border of stable and metastable.
With an emergent Standard Model new global symmetry violations
would occur in higher dimensional operators, 
suppressed by powers of the large scale of emergence
\cite{Jegerlehner:2013cta,Bass:2020gpp}.
Connected to space-time translational invariance,
the cosmological constant scale comes out similar 
to the size of neutrino masses, 
suppressed by power of the large emergence scale.

The plan of this paper is as follows.
We next explain the concept of emergence in particle physics.
Then in Section 3 we discuss the scale hierarchies associated 
with renormalization:
the Higgs mass and zero-point energies of quantum field theory.
Section 4 concerns the full Standard Model
and the role of running masses and couplings 
in understanding 
the particle physics scale hierarchies.
In Section 5 
we discuss the cosmological constant,
where particle physics combines with gravity.
With an emergent Standard Model,
the tiny value of the cosmological constant 
puts an anthropic upper bound on the size of the Higgs mass.
Conclusions are given in Section 6.

\section{Emergence}

Emergence in physics occurs when a many-body system exhibits collective behaviour in the infrared
that is qualitatively different from that of its more primordial constituents as probed in the ultraviolet
\cite{Anderson:1972pca,Kivelson:2016}.
As an everyday example of emergent symmetry, consider a carpet which looks flat and translational invariant when looked at 
from a distance.
Up close, e.g. as perceived by an ant crawling on it, 
the carpet has structure and this translational invariance 
is lost.
The symmetry perceived in the infrared, 
e.g. by someone looking at it from a distance, 
``dissolves'' in the ultraviolet when the carpet is observed
close up.

For emergent particle physics,
the key idea is that for a critical statistical system 
deep in the ultraviolet, close to the Planck scale, 
the only long range correlations 
-- light mass particles --
that might exist in the infrared self-organize into multiplets just as they do in the Standard Model \cite{Jegerlehner:2013cta}.
The vector modes would be the gauge bosons of U(1), SU(2) and SU(3).
In the self-organization process
small gauge groups will most likely be preferred.
Gauge invariance would be exact 
(modulo spontaneous symmetry breaking)
in the energy domain of the infrared effective theory.
Going above the scale of emergence, nature would be described 
by (very possibly) completely different physics with different
degrees of freedom.
Possible emergent gauge symmetries in particle physics were discussed in early work by 
Bjorken \cite{Bjorken:2001pe},
Jegerlehner \cite{Jegerlehner:2013cta,Jegerlehner:1998kt}
and Nielsen and collaborators \cite{Forster:1980dg}.
Recent discussion is given in \cite{Bass:2020gpp,Witten:2017hdv,tHooft:2007nis,Wetterich:2016qee,Jegerlehner:2018zxm}.
Emergent gauge symmetries, 
where we make symmetry instead of breaking it,
are observed in many-body quantum systems 
beyond the underlying QED symmetry and atomic interactions
\cite{Wen:2004ym,Levin:2004js,Zaanen:2011hm}.

With emergence the Standard Model becomes an effective theory
valid up to some large scale, the scale of emergence.
The usual Standard Model action is described by terms of 
mass dimension four or less.
In addition, 
with emergence one also finds an infinite tower of higher 
mass dimensional interaction terms with contributions 
suppressed by powers of a large ultraviolet scale $M$
which characterizes the limit of the effective theory.
If we truncate the theory to include only operator terms 
with mass dimension at most four, then gauge invariant renormalizable interactions strongly constrain 
the global symmetries of the theory which are then inbuilt.
For example, electric charge is conserved and there is 
no term which violates lepton or baryon number conservation.
The dimension-four action describes long distance particle
interactions.
Going beyond mass-dimension four one finds gauge invariant
but non-renormalizable terms
where global symmetries are more relaxed and
which are suppressed by powers of 
the large ultraviolet scale associated with emergence.
Possible lepton number violation, 
also associated with Majorana neutrino masses, 
can enter at mass-dimension five, 
suppressed by a single power of the large emergence scale
\cite{Weinberg:1979sa}.
Baryon number violation can enter at dimension six, 
suppressed by the large emergence scale squared
\cite{Weinberg:1979sa,Wilczek:1979hc}.
Constraints from neutrino masses and proton decay searches 
suggest a scale of emergence 
in the region of $10^{15}$ 
to $10^{16}$ GeV \cite{Bass:2020gpp}.

With emergence, global symmetries would 
be restored with increasing 
large energy 
until we come close to the 
the large energy scale $M$
where higher dimensional terms become important.
Then the system becomes increasingly chaotic with
new global symmetry breaking in the extreme ultraviolet.
This scenario differs from the situation in unification models
which exhibit maximum symmetry in the extreme ultraviolet.

\section{Scale hierarchies in particle physics}

Scale hierarchies arise from the size of QCD and Higgs
condensates compared to the Planck scale as well as
from renormalization effects
involving the Higgs mass and zero-point energies associated
with quantum fields.

The Higgs boson discovered at CERN in 2012 \cite{Aad:2012tfa,Chatrchyan:2012xdj}
completes the particle spectrum of the Standard Model.
In all experimental tests so far it 
behaves very Standard Model like 
\cite{Aad:2019mbh,Sirunyan:2018koj}
and provides masses to the Standard Model particles.

Theoretically, the renormalized Higgs mass squared 
comes with the divergent counterterm
\begin{equation}
m_{h \ {\rm bare}}^2 
= m_{h \ {\rm ren}}^2 + \delta m_h^2
\end{equation}
where
\begin{equation}
\delta m_h^2 
=
\frac{K^2}{16 \pi^2}
\frac{6}{v^2} 
\biggl(
m_h^2 + m_Z^2 + 2 m_W^2 - 4 m_t^2
\biggr)
\end{equation}
relates the renormalized and bare Higgs mass, 
with the renormalized mass connected to the physical 
pole mass.
Here $K$ is an ultraviolet 
scale characterizing the limit to where the Standard Model should work, $v$ is the Higgs vev and the $m_i$ are the Higgs, Z, W and top quark masses. We neglect contributions from lighter mass quarks.
If $K$ is taken as a physical scale, 
then why is the physical Higgs mass so small compared 
to the cut-off?
This is the Higgs mass hierarchy puzzle.
Boson and fermion contributions enter Eq.~(2) 
with different signs.
The renormalized and bare masses would coincide with no hierarchy puzzle if 
\begin{equation}
2 m_W^2 + m_Z^2 + m_h^2 = 4 m_t^2 .
\end{equation}
This equation is the Veltman condition \cite{Veltman:1980mj}.
It implies a collective cancellation between bosons and fermions. Taking the pole masses for the W, Z and top quark (80, 91 and 173 GeV) would require a Higgs mass of 314 GeV, much above the measured value. 
\footnote{Next-to-leading order corrections are suppressed
 by $1/(4 \pi)^2$ and neglected here.}

Pauli \cite{Pauli:1971wp} pointed out that a similar 
situation occurs with the zero-point energies, 
ZPEs, induced by quantization \cite{Bjorken:1965zz}.
Along with condensates associated with spontaneous
symmetry breaking, the ZPEs contribute to the vacuum
energy in particle physics and, 
together with gravitational contributions, to the cosmological constant~\cite{Weinberg:1988cp,Sahni:1999gb,Bass:2011zz}.
Zero-point energies come with ultraviolet divergence requiring regularization and renormalization.
Working in flat space-time
\begin{equation}
\rho_{\rm zpe} 
=
\frac{1}{2} \hbar
\sum_{\rm particles} g_i \int_0^{k_{\rm max}}
\frac{d^3 k}{(2 \pi)^3} \sqrt{k^2 + m^2}
.
\end{equation}
Here 
$m$ is the particle mass;
$g_i = (-1)^{2j} (2j+1) f$
is the degeneracy factor for a particle $i$ of spin $j$, 
with $g_i  >0$ for bosons and $g_i < 0$ for fermions.
The minus sign follows from the Pauli exclusion principle and 
the anti-commutator relations for fermions.
The factor $f$ is 1 for bosons, 
2 for each charged lepton
and 6 for each flavour of quark
(2 charge factors for the quark and antiquark, 
 each with 3 colours).

There is a subtle issue with how to handle ultraviolet divergences consistent with the fundamental symmetries
in the problem.
For example, imagine the example of a two dimensional world 
with circular symmetry.
Then treating divergences involves a circle in momentum 
space extrapolated to infinity.
If we instead sought to use a triangle in momentum space,
the corners and straight edges would violate the underlying 
circular symmetry and might reasonably lead to wrong 
results when connecting to experiments the two dimensional physicist might perform.
A well known example where two classical symmetries clash 
with quantum effects associated with ultraviolet momenta
is the chiral anomaly.
The vector vector axial-vector triangle diagram cannot be
evaluated in a way that preserves gauge invariance (current
conservation) at the vector vertices 
$\gamma_{\alpha}$ and $\gamma_{\beta}$
while preserving chiral
symmetry at the axial-vector vertex $\gamma_{\mu} \gamma_5$. 
Gauge invariance wins with the correction in the axial-vector
current leading to the correct decay rate for 
$\pi^0 \to 2 \gamma$ in QED
\cite{Adler:1969gk,Bell:1969ts}
and the large $\eta'$ mass in QCD \cite{Bass:2018xmz}.

For the ZPEs 
it is important to choose a Lorentz covariant 
regularization procedure
to ensure that the renormalized zero-point energy satisfies 
the correct vacuum equation of state. 
Dimensional regularization with minimal subtraction, 
$\overline{\rm MS}$, is a good regularization.
One finds
\begin{equation}
\rho_{\rm zpe} = - p_{\rm zpe}
=
- 
\hbar \ g_i \
\frac{m^4}{64 \pi^2} 
\biggl[ \frac{2}{\epsilon} + \frac{3}{2} - \gamma
- \ln \biggl( \frac{m^2}{4 \pi \mu^2} \biggr) \biggr]
+ ...
\end{equation}
from particles with mass $m$~\cite{Martin:2012bt}.
Here 
$p_{\rm zpe}$ is the pressure,
$D=4-\epsilon$ the number of dimensions, 
$\mu$ the renormalization scale and $\gamma$ is Euler's constant.
If one instead uses a brute force cut-off on the 
divergent integral, 
the leading term in the ZPE proportional 
to $k_{\rm max}^4$ 
obeys the radiation equation of state $\rho = p/3$.
Eq.~(5) means that the ZPE vanishes for massless
particles, e.g., the photon. 
For Standard Model particles the ZPE
is induced by the Higgs mechanism.

Bosons and fermions contribute to the net zero-point 
energy with different signs. 
This led Pauli to suggest a collective cancelation of the ZPE \cite{Pauli:1971wp}, much like the Veltman condition for the 
Higgs mass squared. 
If we wish to cancel the net ZPE,
then the Pauli equivalent to the Veltman condition reads
\cite{Pauli:1971wp,Kamenshchik:2018ttr}
\begin{eqnarray}
6 m_W^4 + 3 m_Z^4 + m_h^4 &=& 12 m_t^4
\nonumber \\
6 m_W^4 \ln m_W^2 
+ 3 m_Z^4 \ln m_Z^2 
+ m_h^4 \ln m_h^2 &=& 12 m_t^4 \ln m_t^2
\end{eqnarray}
where we again neglect the lighter mass quarks.
For the Standard Model
with the physical W, Z and 
top-quark masses, these two equations would need a 
Higgs mass of about 319 GeV and 311 GeV respectively, 
close to the Veltman value of 314 GeV.
With the Standard Model particle masses, 
the ZPE is negative and fermions dominated.

If we want to cancel the Pauli constraints
we need some extra strength in the boson sector.
A popular candidate for possible extra particles beyond
the Standard Model are
2 Higgs Doublet Models, 2HDMs~\cite{Ivanov:2017dad}.
These are a simple extension of the Standard Model.
One introduces a second Higgs doublet.
There are 5 Higgs bosons, two neutral scalars $h$ and $H$, 
one pseudoscalar $A$
and two charged Higgs states $H^{\pm}$.
Since the 125 GeV Higgs-like scalar discovered at CERN 
in 2012~\cite{Aad:2012tfa,Chatrchyan:2012xdj}
has so far showed no departure from Standard Model predictions, 
it must be assumed 
in any model with extra Higgs states
that 
one of the neutral scalars $h$ 
is a lot like the Standard Model Higgs.

Theoretical constraints on 2HDMs come from
tree level unitarity, 
vacuum stability and requiring perturbative couplings.
In addition, an extra $Z_2$ symmetry is imposed relating
the two Higgs doublets to eliminate unwanted 
flavour changing neutral currents with Yukawa couplings.
This $Z_2$ symmetry may be softly broken (through a mass 
mixing term).

How do 2HDMs affect the Pauli and Veltman conditions?
Possible extra Higgs states are looked for in direct 
searches \cite{Atlas:note,Flechl:2019dtr}.
The parameter space is constrained 
with lower bounds on the masses
from
global electroweak fits \cite{Haller:2018nnx} 
and rare B-decay processes 
\cite{Misiak:2017bgg,Arbey:2017gmh}.
Different model scenarios depend on the fermion to Higgs 
couplings.
The most constrained are type II models with 
600 GeV $< m_{H^{\pm}}$,
530 GeV $< m_A$
and 400 GeV $<m_H$.
\footnote{
Tighter constraints for type II models were claimed in \cite{Chowdhury:2017aav},
viz.
740 GeV $< m_{H^{\pm}}$,
750 GeV $< m_A$
and 700 GeV $<m_H$.
These lower bounds are above the upper bounds 
from tree level unitarity assuming exact 
$Z_2$ symmetry 
(with no mass mixing soft symmetry breaking term),
viz. 
$m_{H^{\pm}} \leq$ 616 GeV,
$m_A \leq$ 711 GeV
and $m_H \leq$ 609 GeV
with $m_h$ taken to be 125 GeV as measured at the LHC \cite{Gorczyca:2011he}.
}
Here one doublet couples to up type quarks and one to down type quarks and leptons. 
Others are the type I fermiophobic model where all fermions couple to just one doublet, lepton specific (one doublet to quarks and one to leptons) and flipped 
(same as type II except leptons couple to the doublet 
 with up type quarks).
There are also inert models where only one doublet acquires a 
vev and couples to fermions.
These models are less well constrained.
For the Veltman condition extended to 2HDMs, 
a favoured benchmark point
is quoted in the type II model 
with $m_H \sim 830$ GeV and $m_A, m_{H^{\pm}} \sim 650$ GeV 
\cite{Darvishi:2017bhf}.
For the mass constraints quoted for the Type II models, 
we would need also extra fermions in the energy range of 
the LHC to cancel the Pauli condition if this scenario is
manifest in nature.

\section{Scale hierarchies with running masses and couplings}

Standard Model particle masses and couplings are related by
\begin{equation}
m_f = y_f \frac{v}{\sqrt{2}}  \ \ \ \ \ 
(f = {\rm quarks \ and \ charged \ leptons})
\end{equation}
where $y_f$ are the Yukawa couplings,
\begin{equation}
m_W^2 = \frac{1}{4} g^2 v^2 , 
\ \ \ 
m_Z^2 = \frac{1}{4} (g^2 + g'^2 )v^2 
\end{equation}
with $g$ and $g'$ the SU(2) and U(1) electroweak couplings,
and
\begin{equation}
m_h^2 = 2 \lambda v^2
\end{equation}
where $\lambda$ is the Higgs self coupling.

The SU(2) and QCD SU(3) couplings, $g$ and $g_s$ are
asymptotically free whereas the U(1) coupling $g'$
is non asymptotically free, rising in the ultraviolet.
(The fine structure constant and its 
 generalizations are defined by $\alpha_i = g_i^2/4 \pi$.)
Running of the Higgs self coupling $\lambda$ 
determines the stability of the electroweak vacuum.
With the Standard Model parameters measured at 
the LHC,
$\lambda$ decreases with 
increasing resolution up to some very large scale.
The sign of the $\beta$-function 
\begin{equation}
\beta_{\lambda} = \mu^2 \frac{d}{d \mu^2} \lambda (\mu^2)
\end{equation}
determines the scale evolution of $\lambda$ 
with $\beta_{\lambda}$
dominated by a large negative top quark Yukawa coupling
contribution (without which the sign of $\beta_{\lambda}$
would be positive). 
QCD interactions of top quarks 
are also essential for keeping the $\beta$-function negative. 
Vacuum stability depends on whether $\lambda$
crosses zero or not deep in the ultraviolet
and involves a delicate balance of Standard Model parameters.

\begin{figure}[t!]  
\centerline
{\includegraphics[width=0.95\textwidth]
{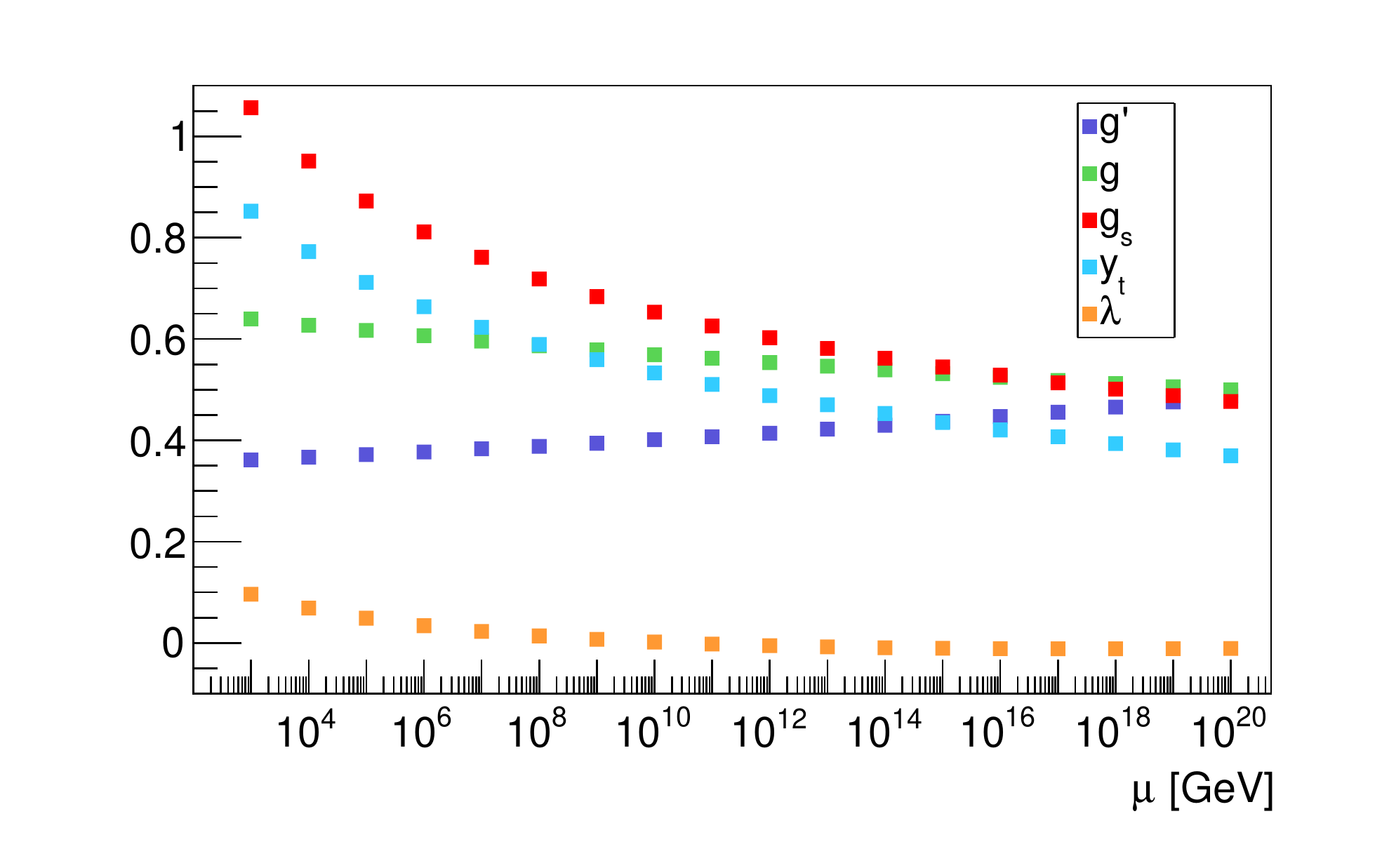}}
\caption{ 
Running of the Standard Model
gauge couplings $g$, $g'$, $g_s$
for the electroweak SU(2) and U(1) and colour SU(3), 
the top quark Yukawa coupling $y_t$ 
and Higgs self-coupling $\lambda$.
(From left, the points describe the evolution of
$g_s$, $y_t$, $g$, $g'$, $\lambda$ in descending order.)
}
\end{figure}

\begin{figure}[t!]  
\centerline
{\includegraphics[width=0.95\textwidth]
{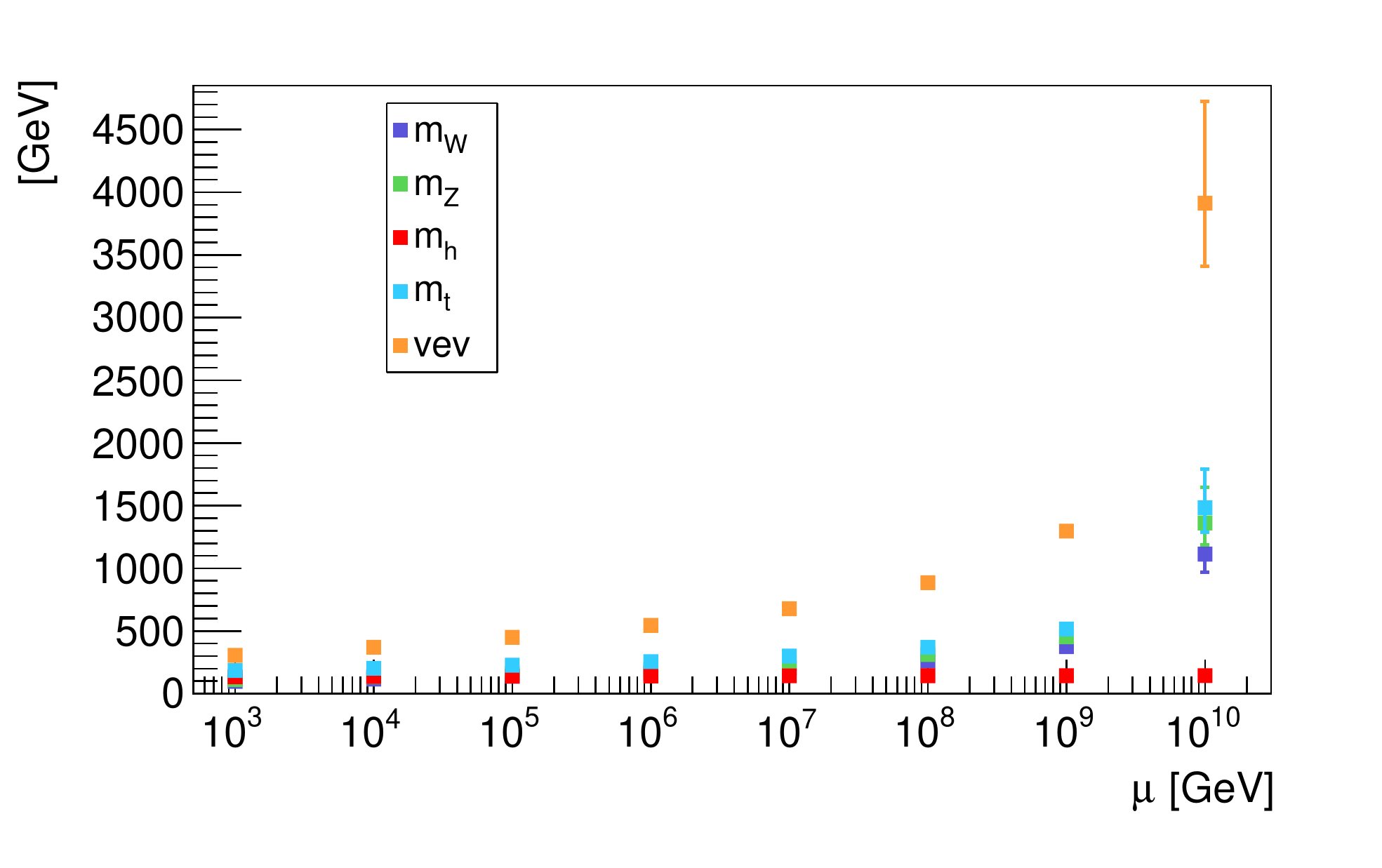}}
\caption{ 
Running $\overline{\rm MS}$ masses and 
the Higgs vev in the Standard Model.
For the relation to the PDG pole masses, 
see~\cite{Kniehl:2016enc}.
Uncertainties are calculated by varying all PDG values
up and down by their respective uncertainties.
(In the printed black and white version,
 the points from top describe the evolution of
 $v$, $m_t$, $m_Z$, $m_W$, $m_h$.)
}
\end{figure}

If we take just the Standard Model with no coupling 
to undiscovered new particles,
then 
one finds that the electroweak vacuum sits 
very close to the border of stable and metastable suggesting possible new critical phenomena in the ultraviolet, 
within 1.3 standard deviations of being stable on 
relating the top quark 
Monte-Carlo and pole masses \cite{Bednyakov:2015sca}. 
\footnote{This $1.3 \sigma$ difference is reduced if 
one includes the difference, about 600 MeV, in the 
top quark and Monte-Carlo
and pole masses discussed in \cite{Butenschoen:2016lpz}.}
Taking the pole mass $m_t = 173$ GeV,
the 125 GeV Higgs mass is close to the minimum needed
for vacuum stability.
If the Standard Model parameters were just
slightly different the low energy effective 
theory emerging from the extreme ultraviolet
would be completely different from the Standard Model
-- see \cite{Bass:2020gpp} and references therein.
The Higgs and other particle masses 
might be linked to physics close to the Planck scale.

Evolution of the Standard Model running couplings is 
shown in Figure 1,
where we evaluate the running couplings 
using the evolution code
mr: Standard Model matching and running C++ package~\cite{Kniehl:2016enc}.
Corresponding to the running couplings in Figure 1,
in Figure 2 we show the running top quark, W, Z and 
Higgs boson masses and the Higgs vev $v$ 
up to the scale, 
just above $10^{10}$ GeV, where $\lambda$ becomes negative
in this calculation with metastable vacuum. 
If here we reduce the PDG top mass to 171 GeV,
then the vacuum stays stable up to the Planck scale.

\begin{figure}[t!]  
\centerline
{\includegraphics[width=0.95\textwidth]
{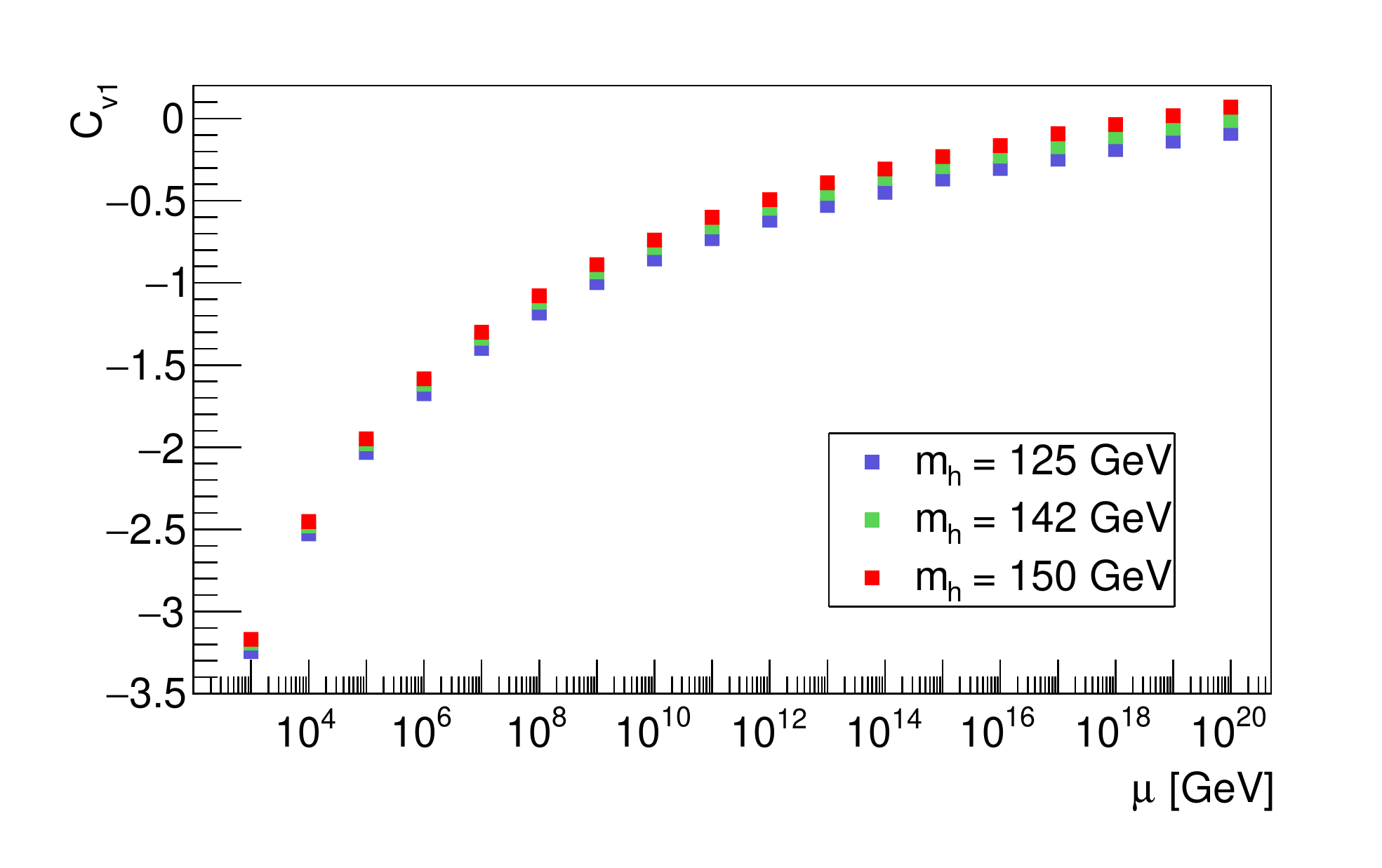}}
\caption{ 
Running of the Veltman coefficient for Standard Model particles.
Here
$
C_{V 1} 
=
\frac{3}{v^2} 
( m_h^2 + m_Z^2 + 2 m_W^2 - 4 m_t^2 )
=
\frac{9}{4} g^4 + \frac{3}{4} g'^4 + 6 \lambda - 6 y_t^2
$
evaluated using the running couplings in Fig.~1.
The points are for Higgs masses $m_h$ equal to 
150, 142 and 125 GeV (top to below).
}
\end{figure}

\begin{figure}[t!]  
\centerline
{\includegraphics[width=0.95\textwidth]
{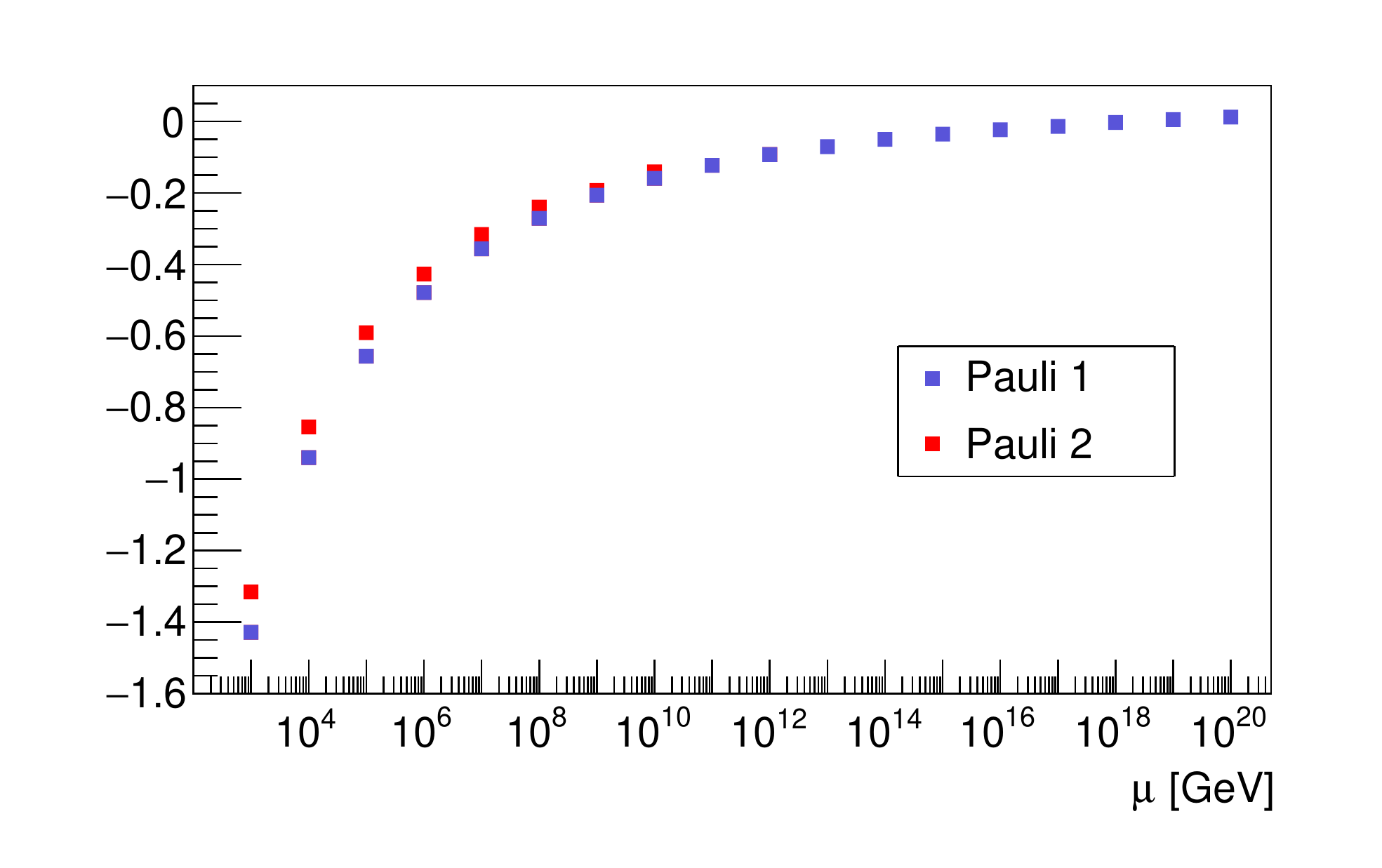}}
\caption{ 
Running values of the Pauli conditions in Eq.~(6)
for Standard Model particles 
(bosons - fermions),
e.g.
Pauli 1 $\{6 m_W^4 + 3 m_Z^4 + m_h^4 - 12 m_t^4 \}$ 
and
Pauli 2
$
\{ 6 m_W^4 \ln m_W^2 
+ 3 m_Z^4 \ln m_Z^2 
+ m_h^4 \ln m_h^2 - 12 m_t^4 \ln m_t^2 \}$.
The Pauli 1 (lower) points are normalized to $v^4$.
The Pauli 2 (upper) points are normalized to $v^4 \ln v^2$
and plotted up to the scale that 
they develop an imaginary part 
when $\lambda$ crosses zero just above $\mu = 10^{10}$ GeV
signaling vacuum instability.
}
\end{figure}

Both the 
Veltman and Pauli constraints are evaluated from 
loop diagrams so the masses which appear there 
are really
renormalization group, RG, scale dependent.
Boson and fermion contributions enter with different signs and evolve differently under RG evolution
which means they have a chance 
to cross zero deep in the ultraviolet.

Veltman crossing means that the renormalized and bare 
Higgs mass squared first coincide,
with the scale hierarchy then radiatively generated
through evolution.
The scale of Veltman crossing is calculation dependent.
With the Standard Model evolution code~\cite{Kniehl:2016enc},
crossing is found at the Planck scale with a Higgs mass about
150 GeV, and not below with the measured mass of 125 GeV
-- see Figure 3 for input PDG masses of 125, 142 and 150 GeV.
If we take input values $m_t=171$ GeV and $m_h=125$ GeV
leading to a stable vacuum in this calculation, then Veltman
crossing happens not below the Planck scale.
In alternative calculations,
Veltman crossing was reported at
$10^{16}$ GeV with a stable vacuum
\cite{Jegerlehner:2013cta}, 
about $10^{20}$ GeV
\cite{Masina:2013wja} 
and much above 
the Planck scale of $1.2 \times 10^{19}$ GeV
\cite{Degrassi:2012ry,Hamada:2012bp} 
with a metastable vacuum.

Figure 4 shows the evolution of the two Pauli constraints
in Eq.~(6), again using the evolution code in 
\cite{Kniehl:2016enc}.
The first
Pauli condition with terms 
$\propto m^4$ crosses zero above $10^{16}$ GeV
corresponding to the net bosonic ZPE contribution outgrowing
the fermionic top quark contribution.
The second Pauli condition is shown up to $10^{10}$ GeV,
above which $\lambda$ becomes negative.
With negative $\lambda$, 
the combination $m_h^4 \ln m_h^2$
develops an imaginary part corresponding 
to vacuum instability;
$[ \ln (-\lambda) = \ln \lambda - i \pi$ for $\lambda > 0 ]$.
For the stable vacuum case with inputs 
$m_t=171$ GeV and $m_h=125$ GeV, 
one finds that both Pauli curves cross zero between 
$10^{17}$ and $10^{18}$ GeV in this calculation.
With a stable vacuum $\lambda$ remains positive definite 
so that $v$ remains finite and the second Pauli condition
develops no imaginary part.

If the Standard Model is emergent below some large ultraviolet
scale $M$,
e.g. associated with vacuum stability and perhaps close
to the scale
where $\lambda$ crosses zero,
then the Standard Model will ``dissolve'' into more primordial
degrees of freedom at this scale.
With an emergent Standard Model,
extrapolating perturbative evolution calculations above 
any scale of emergence corresponds to extrapolating into an unphysical region since the degrees of freedom there will be completely different.

\section{Vacuum energy and the cosmological constant}

Vacuum energy is measured through the cosmological constant
$\Lambda$
which appears in Einstein's equations of General Relativity.
Before we couple to gravity only energy differences 
have physical meaning,
which 
allows us to cancel the ZPE through normal ordering.

Einstein's equations read
\begin{equation}
R_{\mu \nu} - \frac{1}{2} g_{\mu \nu} \ R = 
- \frac{8 \pi G}{c^2} T_{\mu \nu} + \Lambda g_{\mu \nu} .
\end{equation}
Here $R_{\mu \nu}$ is the Ricci tensor, 
$R$ is the Ricci scalar
and 
$T_{\mu \nu}$ is the energy-momentum tensor 
for excitations above the vacuum;
$G$ is Newton's constant and $c$ is the speed of light.
These equations determine the geodesics on which particles
propagate in curved space-time.
The cosmological constant
measures the vacuum energy density
\begin{equation}
\rho_{\rm vac} = \Lambda / (8 \pi G) .
\end{equation}
It receives contributions 
from the ZPEs,
any (dynamically generated) potential in the vacuum,
e.g. induced by the QCD and Higgs condensates, 
and a renormalized version of 
the bare gravitational term $\rho_{\Lambda}$
~\cite{Sola:2013gha}
\footnote{Note that $\rho_{\rm \Lambda}$ 
corresponds to $V_0$ Eq.~(3.8) of \cite{Weinberg:1988cp}.}, 
viz.
\begin{equation}
\rho_{\rm vac} = \rho_{\rm zpe} + \rho_{\rm potential} 
+ \rho_{\rm \Lambda}.
\end{equation}

Matter 
clumps together under normal gravitational
attraction whereas the cosmological constant 
is the same at all points in space-time and
drives the accelerating expansion of the Universe.
As an observable 
the cosmological constant
is renormalization scale invariant.
It is
independent of how a theoretician might choose to calculate it\footnote{Here General Relativity is taken as a classical theory 
 with Newton's constant RG scale invariant.},
\begin{equation}
\frac{d}{d \mu^2} \rho_{\rm vac} =0 . 
\end{equation}
On distance scales much larger than the 
galaxy the Universe exhibits a large distance flat geometry. 
Observations based on supernovae type 1a, 
the large scale distribution of galaxies and 
the Cosmic Microwave Background 
\cite{Aghanim:2018eyx,Frieman:2008sn}
point to a small positive value for the 
cosmological constant corresponding to 
\begin{equation}
\rho_{\rm vac} = (0.002 {\rm \ eV})^4
\end{equation}
and a present period of 
accelerating expansion 
that began about five billion years ago.

Historically, Einstein introduced the cosmological constant
in an attempt to give a static Universe \cite{Einstein:1917}.
Shortly afterwards, he expressed doubts describing $\Lambda$
as 
``greatly detrimental to the formal beauty of the theory''
\cite{Einstein:1919}.
The static Universe solution proved unstable to local inhomogeneities in the matter density. 
Einstein abandoned the cosmological constant,
setting it equal to zero, 
following Hubble's observation of an expanding Universe
\cite{Einstein:1931}.
Feynman in his lectures on gravitation also wrote that 
he believed 
Einstein's second guess and expected a zero cosmological
constant \cite{Feynman:1996kb}.
It returned to physics with discovery of the accelerating
expansion of the Universe.

Whereas the total
$\rho_{\rm vac}$ is renormalization scale invariant,
individual contributions in Eq.~(13) do carry scale
dependence.
For example, 
the ZPE contributions in Eq.(5) are scale dependent
both through explicit $\mu^2$ dependence and through
the running masses.
The Higgs potential is RG scale dependent 
through 
the scale dependence of 
the Higgs mass and Higgs self-coupling,
which determines the stability of the electroweak vacuum.
This renormalization scale dependence 
cancels to give the scale invariant $\rho_{\rm vac}$.
The important question is whether there is anything left over.
How big is the remaining $\rho_{\rm vac}$?
How do we understand the measured tiny value in Eq.~(15)
with scale 0.002 eV when individual contributions involve
the QCD and electroweak scales?

One finds a simple explanation with an emergent
Standard Model.
With a finite cosmological constant Einstein's equations 
have no solution 
where $g_{\mu \nu}$ is the constant Minkowski metric \cite{Weinberg:1988cp}.
That is,
space-time translational invariance 
(a subgroup of the group of 
 general co-ordinate transformations)
is broken without extra fine tuning. 
The reason is that
$\rho_{\rm vac}$ acts as a gravitational source 
 which generates a dynamical space-time, 
 with accelerating
 expansion for positive $\rho_{\rm vac}$.
 (For a Universe dominated by the cosmological constant
  space-time is 
  described by the de Sitter metric, 
  $ d s^2 = d t^2 - e^{2 H_{\infty} t}
    (d r^2 + r^2 d \theta^2 + r^2 \sin^2 \theta d \phi^2)$
  where $H_{\infty}^2 = \frac{1}{3} \Lambda$ is the Hubble
  constant in the infinite future.)
A large net $\rho_{\rm vac}$ would challenge 
the successful phenomenology of 
Special Relativity and particle physics with flat space-time
in our experiments.

With the Standard Model as an effective theory emerging 
in the infrared,
low-energy global symmetries can 
be broken through 
additional higher dimensional terms,
suppressed by powers of the
large emergence scale~\cite{Witten:2017hdv}. 
Suppose the vacuum including condensates with finite vevs
is translational invariant and 
flat space-time is consistent at dimension four, 
just as suggested by the success of the Standard Model
and Special Relativity.
Then 
the RG invariant scales
$\Lambda_{\rm qcd}$ and 
electroweak $\Lambda_{\rm ew}$
might enter the cosmological constant
with the scale of the leading term
suppressed by 
$\Lambda_{\rm ew}/M$, where $M$ is the scale of emergence
(that is, $\rho_{\rm vac} \sim (\Lambda_{\rm ew}^2/M)^4$
 with one factor of 
 $\Lambda_{\rm ew}^2/M$ for each dimension of space-time). 
This scenario, if manifest in nature, 
would explain why the cosmological constant scale 0.002 eV 
is similar 
to what we expect for the neutrino masses 
\cite{Altarelli:2004cp},
which for Majorana neutrinos are
themselves linked to a dimension five operator with
$m_{\nu} \sim \Lambda_{\rm ew}^2/M$ \cite{Weinberg:1979sa}.
The cosmological constant would vanish at dimension four.
In this sense Einstein's second guess, 
also Feynman's
guess, would be correct:
the cosmological constant vanishes if we truncate 
the action to terms of mass dimension four or less.
This vanishing cosmological constant is equivalent
to a renormalization condition $\rho_{\rm vac} =0$
at dimension four imposed by global space-time translational
invariance, 
even in the presence of large QCD and Higgs condensates.
The precision of global symmetries in our experiments,
e.g. 
lepton and baryon number conservation, 
tells us that the scale of emergence should 
be deep in the ultraviolet, 
much above the Higgs and other Standard Model particle masses. 
Taking 
$0.002 {\rm \ eV} = \Lambda_{\rm ew}^2/M$ 
gives a value of $M$ about $10^{16}$ GeV.

The tiny cosmological constant enters as a subleading
term in the low-energy expansion of the action for the
emergent Standard Model.
Within this scenario anthropic arguments place an upper
bound on the value of $\Lambda_{\rm ew}$.
It is interesting that the parameters of particle physics
interactions are fine tuned to our existence \cite{Carr:1979sg,Rees:2005}.
Small changes in particle masses and couplings would lead to 
a very different Universe, assuming that the vacuum remained
stable, 
with one example that 
small changes in the light-quark masses can prevent
Big Bang nucleosynthesis.
Accelerating expansion of the Universe takes over when
the energy density associated with the cosmological constant 
exceeds the mean matter density 
(including dark matter contributions).
Weinberg argued
that if the cosmological constant were 
ten times larger 
the present period of acceleration would have begun 
earlier enough that galaxies would have no time to form
\cite{Weinberg:1987dv}.
With
$\rho_{\rm vac} \sim (\Lambda_{\rm ew}^2/M)^4$
this constraint corresponds
to a factor of 1.33 on $\Lambda_{\rm ew}$ or upper bound 
on the Higgs mass,
which is complementary to the lower bound, about 125 GeV,
needed for electroweak vacuum (meta)stability with other 
PDG parameters held fixed.

\section{Conclusions}

With the great success of the Standard Model at the LHC 
and in low-energy precision experiments, 
it is worthwhile to re-evaluate our ideas about 
the origins of gauge symmetry in particle physics.
Might the gauge symmetries be emergent?
The (meta)stability of the electroweak vacuum suggests 
that the Standard Model parameters measured in 
experiments
might be correlated with physics deep in the ultraviolet.
Global space-time translational symmetry and 
the successful phenomenology of flat space-time in 
laboratory experiments and our everyday experience
is consistent with emergent symmetry,
with the cosmological constant scale suppressed by 
power of the large scale of emergence.
In this scenario the cosmological constant scale
would be similar to the size of Majorana neutrino masses.
The tiny cosmological constant may be teaching us about 
the deeper origin of symmetry in particle physics.
Future experiments will measure the dark energy equation
of state with the EUCLID mission of ESA expected 
to be sensitive to variations from a time-independent 
cosmological constant of 10\% or more~\cite{Laureijs:2011gra}.
Next generation neutrinoless double $\beta$-decay
experiments~\cite{Agostini:2017jim,Caldwell:2017mqu},
e.g.
the future LEGEND 1000 tonne experiment at Gran Sasso
\cite{Abgrall:2017syy},
will be sensitive to 
Majorana neutrinos with mass in the range
of the scale of the cosmological constant scale, 0.002 eV.

\bibliographystyle{unsrt}

\end{document}